\documentclass{egpubl}
\usepackage{EGUKCGVC2021}
\usepackage{xcolor}

\WsPaper           % uncomment for final version of EG Workshop contribution
\usepackage[T1]{fontenc}
\usepackage{dfadobe}  

\usepackage{cite}  % comment out for biblatex with backend=biber
\BibtexOrBiblatex
\PrintedOrElectronic
\ifpdf \usepackage[pdftex]{graphicx} \pdfcompresslevel=9
\else \usepackage[dvips]{graphicx} \fi

\usepackage{egweblnk} 
\graphicspath{{pictures/}}

\usepackage{enumitem}
\usepackage{caption}
\usepackage{subcaption}
\usepackage{adjustbox}

\usepackage[T1]{fontenc}
\title[Learning Activities in Colours and Rainbows for Programming Skill Development]%
      {Learning Activities in Colours and Rainbows\\for Programming Skill Development}
  
\author[J.\,C. Roberts]
{\parbox{\textwidth}{\centering J.\,C. Roberts\thanks{Email j.c.roberts@bangor.ac.uk}\orcid{0000-0001-7718-3181}}\\
{\parbox{\textwidth}{\centering Bangor University, UK\vspace{-5mm}}
}
}

\begin{document}

% uncomment for using teaser
% \teaser{
%  \includegraphics[width=\linewidth]{eg_new}
%  \centering
%   \caption{New EG Logo}
% \label{fig:teaser}
%}

\maketitle
%-------------------------------------------------------------------------
\begin{abstract}
We present how we have created a series of bilingual (English and Welsh) STEM activities focusing on rainbows, colours, light and optical effects. The activities were motivated by the many rainbows that appeared in windows in the UK, in support of the National Health Service at the start of the coronavirus pandemic. Rainbows are hopeful and are very fitting to be used as a positive iconic image at a time of much uncertainty. In this paper we explain how we have developed and organised the activities, focusing on colours, computer graphics and computer programming. Each lesson contains one or more activities, which enable people to take an active role in their learning. We have carefully prepared and organised several processes to guide academic colleagues to create and publish different activities in the theme. Which means that the activities appear similarly structured, can be categorised and searched in a consistent way. This structure can act as a blueprint for others to follow and apply to develop their own online course. The activities incrementally take people through learning about colour, rainbows, planning what to program, design and strategies to create colourful pictures using simple computer graphics principles based in processing.org. 
%-------------------------------------------------------------------------
%  ACM CCS 1998
%  (see https://www.acm.org/publications/computing-classification-system/1998)
% \begin{classification} % according to https://www.acm.org/publications/computing-classification-system/1998
% \CCScat{Computer Graphics}{I.3.3}{Picture/Image Generation}{Line and curve generation}
% \end{classification}
%-------------------------------------------------------------------------
%  ACM CCS 2012
%The tool at \url{http://dl.acm.org/ccs.cfm} can be used to generate
% CCS codes.
%Example:
\end{abstract}  
%-------------------------------------------------------------------------
%\setlength{\parskip}{2mm plus2mm minus2mm}
\section{Introduction}
During the start of the coronavirus pandemic, rainbows appeared in many houses in the UK in support for the National Health Service. Rainbows represent hope, and it seemed fitting that people created colourful images at a time of much uncertainty. Indeed, as academics who teach computer graphics~\cite{roberts2021visualisation}, we wanted to help people. Being educators, we started to consider what we could teach. Especially, during the early lockdown in the UK,
it was clear that students had an appetite to learn, but were not sure what material to follow. Through seeing many rainbows in people's front windows, we started to think that we could create a set of learning activities around colours and rainbows, and make them publicly available. The activities would perhaps excite and encourage people to learn some basic programming skills. The idea was born, and we called it ``Project Rainbow''~\cite{ProjectRainbow2021}. The plan was to create a series of lessons, around the idea of colours, light and rainbows. Our focus was to develop content for people who wish to learn computer programming and visual computing skills; students at secondary school, college or universities who wish to follow an introductory to coding course. We would get learners to consider how rainbows are formed in the natural world and how rainbow pictures can be created in code.

\vspace{-1mm}
In this paper we describe the structure of our short course, and give examples of the lessons. Throughout the paper we highlight specific points that we want learners to understand, and skills that we want them to develop. These include to critically think about the work before coding (and not to just jump into programming); to learn more about colours; develop basic computer graphics skills (such as using the painters algorithm, coding colours on a computer~\cite{foley1996computer}); learn and develop their coding skills; and above all to have fun with STEM activities around colours and graphics. We focus on learning activities and developed our course in WordPress, which allowed several academics to create and post activities and have them peer moderated before final publication. We insisted that each lesson was a \textit{learning activity} and include something that the learner should achieve. This was important, because we wanted the learners to actively partake in the exercises. We also defined a structure for the activities (criteria of how the lessons should be written and structured), which allowed different academics to add their own lessons. It meant that each of the activities followed a similar structure, were tagged appropriately (so that they can be searched) and had a similar appearance.  

\vspace{-1mm}
We organise the paper as follows: after the related work (Section~\ref{sec:relatedwork}) we explain the structure of each learning activity (Section~\ref{sec:pathway}) and the learning pathways, and then explain the material in the suggested learning-pathway order. Subsequently, we cover introduction to colours and the coding environment (Section~\ref{sec:intro}); methods to plan your code and design a scene to be programmed (Section~\ref{sec:critical}); learning about colour models, and creating colourful patterns (Section~\ref{sec:colour}); and finally advancing beyond basic Java to create an augmented reality rainbow and one in pure CSS (Section~\ref{sec:advanced}). In this work we contribute a careful structure of the learning activities and principles, that can be followed by other people who wish to develop learning materials online. We also contribute a set of lessons around colours and rainbows, that learners may wish to follow, or teachers use in their own lesson planning.

\vspace{-1mm}
\section{\label{sec:relatedwork}Related work}
As teachers we are keen to encourage \textit{active learning} strategies. Where students take an active role in their own learning. With Project Rainbow we knew we would put the information on a website, with online material, include activities for students to follow, and that students would work remotely and on their own.  Consequently, we needed to include strategies to help people take an active role in their own learning. The material should lead them through the concepts, get them to create something, and then allow them to reflect on what they are learning~\cite{bonwell1991active}.  In fact, through most of our teaching, we are keen to encourage experiential learning~\cite{KolbKolb2005}. Where the learner can develop their own skills in design, creative thinking and crafting of computer graphics pictures~\cite{Roberts_ETAL2018_EVF}. Consequently, in Project Rainbow we decided to encourage a sequential approach, where students work through a series of activities, and mix together background and theoretical information with practical skills. Each of the exercises are designed to help learners analyse and synthesise the material, and then reflect on their work~\cite{krathwohl2002revision}. They are given additional questions at the end of each activity, to further their knowledge. Additionally, through the material we add links to help them relate the information to other resources that are external to our learning site. In this way, we hope to develop not only intellectual knowledge, but also allow learners to practice that knowledge~\cite{kennedy1999role}. 

To organise this material, there are several different structures and frameworks that we could follow. For instance, the decision processing model by Simon~\cite{Simon1973structure} suggests splitting the work into Intelligence, Design, Choice, Implementation and review. However we wanted to make each part stand-alone and this model relies on teachers explaining much background information first. We could follow our Explanatory Visualisation Framework (EVF)~\cite{Roberts_ETAL2018_EVF} process, which encourages students to perform their own research on a topic, create a report, make a design, plan, develop and then reflect. But the EVF structure also requires that the teacher intervenes at specific parts, to evaluate the research report, design document and so on, which did not seem appropriate for our goals. However the structure of the EVF and those by Munzner~\cite{Munzner2009}, McKenna et al. \cite{McKennaETAL2014} and Sedlmair et al. \cite{Sed2012a} have similar traits. Each elaborate around four key stages: performing research, design, development and reflection. We chose to adopt this structure. But rather than asking the learners to perform their own research on the topic, we instead explain the background and reference material. We include references to material on the web, for them to follow and reference, so that they can do their own research. Students learn about the ideas (research), they consider and envision what the visual output would look like (design), achieve the activity (development), and finally reflect on what they have done.   

Critical thinking is one of the important lessons we want students to learn~\cite{alsaleh2020teaching}. We believe it is important that learners make plans, have a clear vision of what they want to create and critically think throughout their activity. We have observed that learners have a tendency to jump straight into coding, without really thinking or planning what they are doing~\cite{Roberts-et-al-PDVW-2017}. Consequently, there is much worth in getting learners to first reflect and think-critically on what they are about to do~\cite{RobertsHeadleandRitsos2017}. We do not want students to write all their code on paper, however we do want them to understand what they want to achieve with their code. Drawing on our previous experience in this area, we could ask the learners to perform a Five Design-Sheet (FdS) study of their ideas~\cite{RobertsHeadleandRitsos15,RobertsHeadleandRitsos2017}. However, the full FdS sketched design-study can take students over two hours to complete, and therefore we felt that it would be too time consuming, and the early learners could either not complete it or perhaps be discouraged through trying to achieve it. In addition, we wanted learners to critically consider what they were going to achieve. Through the activity they should  try to realise the main coding principles that they would need and ascertain if they needed to do any more self-study in coding. Subsequently, we decided to follow the Critical Thinking Sheet method (CTS)~\cite{roberts2019critical,RobertsRitsos_CTS_2020}. This strategy is quick to perform, as learners only need to complete one page, and it provides a picture of what they will create.

We acknowledge that there are other activities that we could have included, such as brainstorming using cards (e.g., vizitcards~\cite{HeAdarVisitCards}),
construction by using tokens~\cite{huron2014constructiveviz}, get the students to do some informal sketching~\cite{buxton2010sketching}, or perhaps look do their own research and make a table of tools~\cite{Ridley_ETAL_VisToolsActivity2020}, which we could adapt to be a table of colour tools. However, we chose to develop a new set of activities. We wanted the activities to directly relate to each lesson, be focused on rainbows and colours, and address particular learning goals of that activity. Additionally, to break up the coding challenges, we decided to also add some crafting activities and encourage learners to craft something~\cite{Lowgren2016}, without the need of a computer, as per the principles of CS unplugged activities~\cite{bell2018cs}. 

\begin{figure*}[t]
    \centering
    \includegraphics[width=\textwidth]{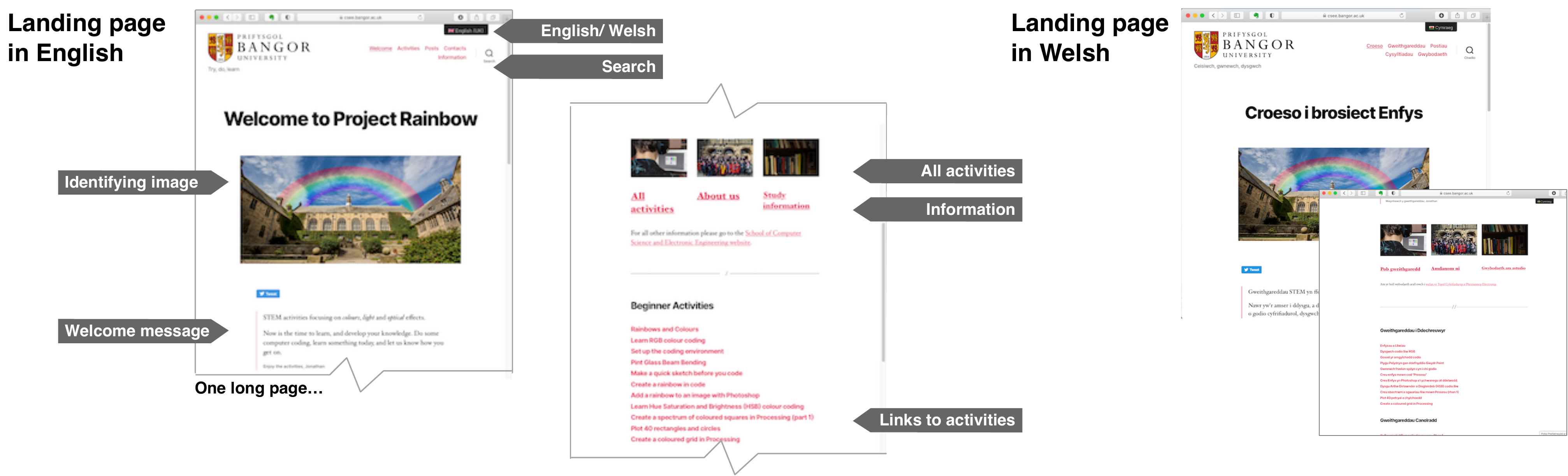}\vspace{-2mm}
    \caption{Landing page of Project Rainbow~\cite{ProjectRainbow2021}. Showing the WordPress setup, menu to change between English and Welsh, identifying image, welcome message, and links to activities.}
    \label{fig:landingpage}
\end{figure*}

Finally, while rainbows are definitely colourful, and can be used to create interesting computer graphics pictures, there can be several challenges when using rainbows. These issues manifest when using rainbow colours to display data, represent information, or present information consistently across devices. One principal challenge is that the rainbow colour map is not continuous in illuminance~\cite{Ware2012}. Therefore, data displayed in a rainbow colour map can mislead users~\cite{borkin2011}. Another challenge is that sometimes low and high values are mapped to dark colours (at the extremes of the rainbow hue), which can lead to misunderstanding of the data. In fact, many researchers have campaigned for people to not use the rainbow colour map. Rogowitz and Treinish call for `the end of the rainbow'~\cite{RogowitzTreinish1998,RogowitzKalvin2001_WhichBlair}; Borland and Taylor suggest that the rainbow colour map is still harmful~\cite{borland2007rainbow}, while Crameri et al.\ \cite{CrameriETAL2020_MisuseOfColour} explain wider issues of colour in science communication.
While we acknowledge these challenges when using the rainbow colour map, our goals are different -- to create a series of interesting learning activities around colours.

\section{\label{sec:pathway}Developing the activity structure and learning pathways}
To develop the site we decided to use WordPress (WP). We made this decision because we wanted to have an editing permissions infrastructure to allow different people to edit specific parts of the site, and it would be mobile ready, accessible and available on different platforms. We developed three parts to the learning site: landing page and information, activities and study guide. We wanted learners to locate the activities as quickly as possible. The landing information are created as \textit{Pages} in WordPress, that follow the main template design. Activities are WordPress \textit{Posts}, and the study guide is a dynamic page built from tagged posts.

Figure~\ref{fig:landingpage} shows our \textbf{landing page}, where users select the activities. Only specific authors have permission to edit these pages. We chose a clean-looking style, branded the pages with the University logo and followed the University's style guide. We wanted to keep minimal information on the front page, explain that it was ``STEM activities focusing on colours, light and optical effects'', include the list of activities, and provide a search for interesting activities. WordPress also allows different plugins to extend its main functionality. We used two additional plugins: first a translation plugin to allow us to have dual language support (English and Welsh). Most academics started writing their activity in English, which was translated into Welsh by themselves, and/or checked by a Welsh speaker. Second, we wanted to keep logs of activities to understand how the site was being used, and so needed to use cookies. Subsequently we developed a privacy notice and  cookie policy, and allow users to accept or reject our use of cookies. 

\begin{figure*}
    \centering
    \includegraphics[width=\textwidth]{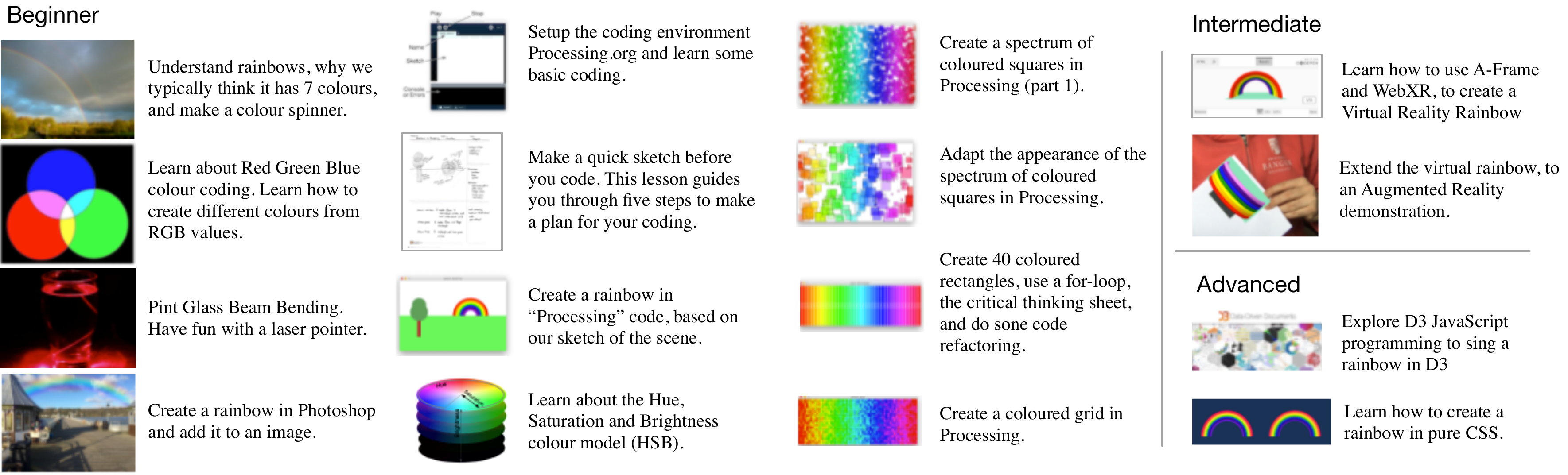}
    \caption{We have created several different lessons. Each is presented to the student with an indicative picture, short description and links on the website to the main activity. This information is displayed on the learning-pathway, with appropriate headings (e.g., Colour, planning, patterns) and starts with beginner, then intermediate and ends with advanced activities.}
    \label{fig:learningpathway}
\end{figure*}

Using WP meant that we could create individual \textbf{activities} as blog posts, post them to the site when they were available. Using the permissions and editing capability of WP meant that several academics could create the activities, and not have access to the wider setup of the site. We also developed a set of written guidelines over how to write, create, categorise and tag each post. Consequently, each post would look similar, have allocated \textit{tags} to allow search-engines to find individual activities, and additional \textit{categories} to organise the material on the site. Table~\ref{tab:tags} shows two example activities. We needed to develop guidelines for authors, because, while WP provides the ability to create categories and tags, it does not provide a list of the available ones. Consequently we included five parts to the guidelines. 

\begin{table}[t]
\centering
  \caption{Two example activities with title, categories and tags. We gave guidelines to academics to create appropriate descriptive titles, add categories and especially the level of Beginner, Intermediate or Advanced, and a set of descriptive tags.}
  \flushleft
  \label{tab:tags}%
\begin{flushleft}
  \resizebox{\columnwidth}{!}{%
  \renewcommand{\arraystretch}{1.5}
  \hspace*{-4mm}
  \begin{tabular}{@{}p{3cm}p{3.5cm}p{3.5cm}@{}}
  \textbf{Title}&\textbf{Categories}&\textbf{Tags}\\\hline
  Create a coloured grid in Processing&Beginner, Colours, Java, Processing, Project Rainbow& cell, colour, for-loop, grid, hsb, hue, index, jitter, random, saturation, value\\[5mm]
  Plot 40 rectangles and circles& Beginner, Colours, Processing, Project Rainbow& colour, critical thinking sheet, sheet, ellipse, hsb, hue, java, rectangles, refactor, saturation, value\\
    \end{tabular}
    }
    \end{flushleft}
\end{table}

\begin{enumerate}[noitemsep,nolistsep]
    \item Each activity needed a descriptive and short title. The title needs to be short and explain what was to be done by the activity. 
    \item Authors need to categorise each post as ``Project Rainbow'' and label each as Beginner, Intermediate or Advanced. As a group of academics, we deliberated over these categories. They provide a convenient classification of the information. Beginners, we describe as new to programming and do not know much about colour. Intermediate activities, develop skills for students who have coded before, and know something about colour. While the categories labelled advanced are for those who have done a lot of programming already. 
    \item Authors should add five or more tags. The tags must describe different parts to the activity. For example, with the coding tasks, the tags would describe different aspect of the code, such as colour, grid, for-loop, array and so on. They can be then used by search engines to match to the best learning activity.
    \item We provide a draft structure that each authors should follow. Each post should have an introduction to the material, background information with links to external sources and an activity. The activity should be central to the post and if the activity had several parts then it can be mixed with background information, and labelled with several steps. Images or photographs should be used within the material, and have alternative text associated with them. Several ``top tip'' summary boxes should be added, which highlight something to remember. The post must also include a summary description and an indicative picture, which will be used in the learning-pathway section. We also defined the coding style, that we defined in CSS and added to WordPress. This meant that the code fragments appeared the same in different posts. Finally, authors need to organise the translation into Welsh.
    \item Finally each activity is reviewed by a colleague, for its accurateness, quality of post, clarity of instruction and appropriateness to Project Rainbow. Academics could nominate another user of the system to be their reviewer.
\end{enumerate}

\vspace{-1mm}
The final section to the site is the \textbf{study-guide}, screenshots of which are shown in Figure~\ref{fig:learningpathway}. While all learning is a personal experience~\cite{tobin2000all}, and some students will find their own learning strategies~\cite{keppell2014personalised}, we wanted to publish a suggested way to navigate the material. We achieved this by creating a suggested learning-pathway. This is created as a summarised set of blog posts, ordered through beginner, intermediate and advanced and leading the learner through the ideas in a pre-defined way. Students can use this list as an index into the material. We mix  the type of activities in this learning-path, some activities are practical activities without a computer (sketching and building), many involve computer programming using processing.org, and others use JavaScript and CSS. We chose processing.org~\cite{reas2007processing} because it was developed to create visual images, and has been used with learners with wide abilities.  If someone wishes to follow a more ad hoc order, then we help this strategy too. Learners may find specific posts through search-engines, which is helped through tagging and classifying the posts. Or they can search the Project Rainbow site for specific keywords, or select only (for example) \textit{beginner} exercises.

\section{\label{sec:intro} Introductory activities -- learning about colours}
We include several introductory activities. These cover basic information, from how rainbows are formed in nature, how colour can be coded on a computer using RGB colour codes, how lights travel through transparent objects, how to create a rainbow in Photoshop, and setting up the processing.org coding environment. The aim is that by the end of these activities the learner will know something about colours, achieved some practical activities, and have setup the programming environment ready for activities that follow.

When developing introductory learning materials, it can be challenging to know the level and writing style. We did not want to make the information patronising, however we did want to cover basic concepts so that people would be able to do the more complex activities. In addition, we wanted to make the information both accessible and interesting. Subsequently, we mixed basic knowledge, with information that people may not know. For instance, when introducing colours we obviously made reference to the colours of the rainbow (Red, Orange, Yellow, Green, Blue, Indigo and Violet) and made reference to the mnemonic Roy G. Biv. Indeed, this mnemonic is useful because it highlights the main three primary colours \textbf{R}ed, \textbf{G}reen and \textbf{B}lue which helps us lead onto the RGB colours that are used on computers~\cite{foley1996computer}.  We briefly explain the main introductory activities below.

\begin{figure}[b]
    \centering
    \includegraphics[width=\columnwidth]{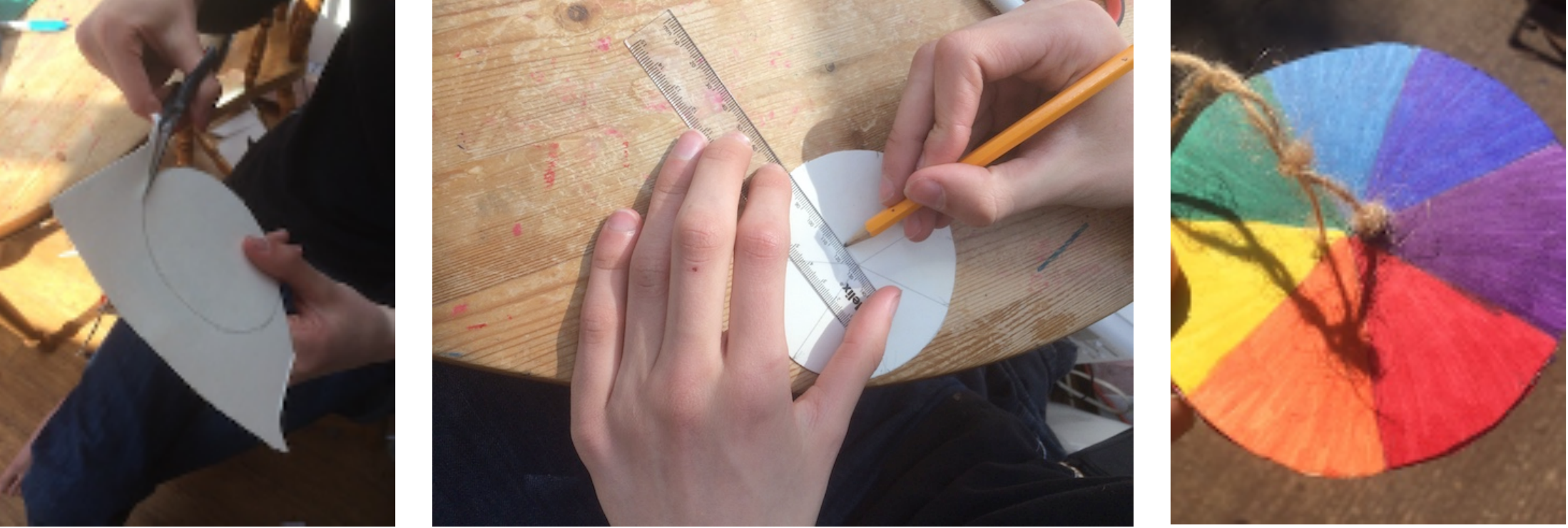}
    \caption{Photographs from the colour spinner activity.}
    \label{fig:colour-spinner}
\end{figure}

The first activity focused around \textbf{rainbows in nature and light} and we led students to create a colour spinner (see Figure~\ref{fig:colour-spinner}).  We explained how rainbows in nature are formed, noticing how the second arc of a double rainbow appear reversed. That the summary of the colours of a rainbow, into seven distinct colours, is attributed to Aristotle (384–322 BC) but it may have been developed by Aristotle’s successors Theophrastus or Strato. This helps to get people to critically think about colours, and understand that the seven colours in a rainbow are artificially named. Indeed, we go on to explain that Newton, with his colour circle, described how coloured lights can be mixed together, and that he only used five colours. But added orange and then indigo to match with the seven notes in a major scale. In addition, giving links to external information, such as Wikipedia pages, helps to allow people read around the topic. 
We discussed mixing paints verses mixing lights (subtractive and additive colour mixing) and defined terms such as colour hues. For the activity we focused on creating a physical colour-wheel, and gave step-by-step instructions how to create and use one.

\begin{figure}[h]
    \centering
    \includegraphics[width=.49\columnwidth]{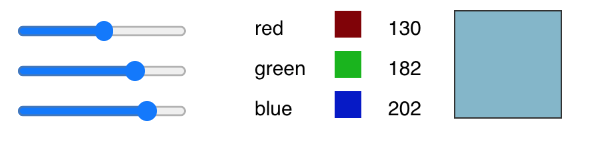}
    \includegraphics[width=.5\columnwidth]{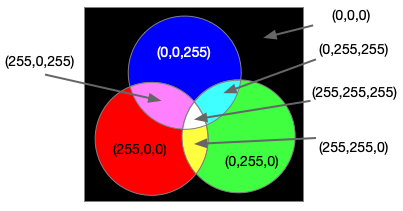}
    \caption{Left shows a screenshot of the colourpicker app, that we embedded into WordPress, to allow people to explore RGB colours. Right we get learners to move between the colours, e.g., starting with magenta and adding green to make it lighter.}
    \label{fig:colour-picker}
\end{figure}

The second activity focused on \textbf{coding colours} on a computer. As a team we had a discussion whether we should teach RGB or Hue-saturation-value (HSV) colour spaces first. HSV has the advantage that the colours are ordered by the rainbow hue; however after deliberation we decided that it was important that learners understand RGB colour coding. In addition, people can find RGB colours confusing to manipulate, and thought it would make a good activity. We were careful in our wording around this topic; to define technical terms when they were needed. We added a further post at a later time, around the topic of colour spaces. There are many colour pickers that could help to explain the RGB colour space (such as Google's colour picker or the W3Schools HTML colour picker). While each have their advantages, we decided to embed our own colour application into the page (created with p5.js), as shown in Figure~\ref{fig:colour-picker} (left). By embedding the application on the page it kept learners on the page, and allowed people to explore different colours. If people wished to look at more advanced colour pickers they could visit the external sites. In addition, it would have been possible to perform a similar activity in processing.org, edit p5.js or in HTML, and we had plans to create activities around these different pickers. However, at this stage, we wanted to focus on basic principles and not rely on coding knowledge, so chose to use the embedded colour picker code and focus on the colours themselves not any specific picking tool.

\begin{figure}[t]
    \centering
    \includegraphics[width=\columnwidth]{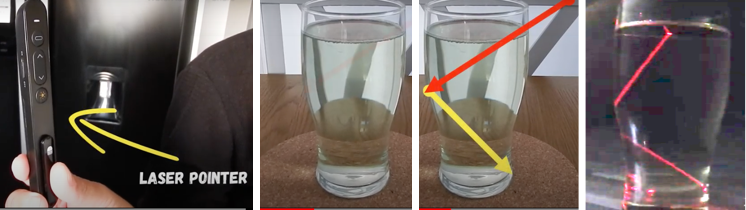}
    \caption{This activity shows how laser light can be made visible in a glass of water with washing liquid, and shows how laser light reflects in the glass.}
    \label{fig:laser}
\end{figure}

\begin{figure}[t]
    \centering
    \includegraphics[width=\columnwidth]{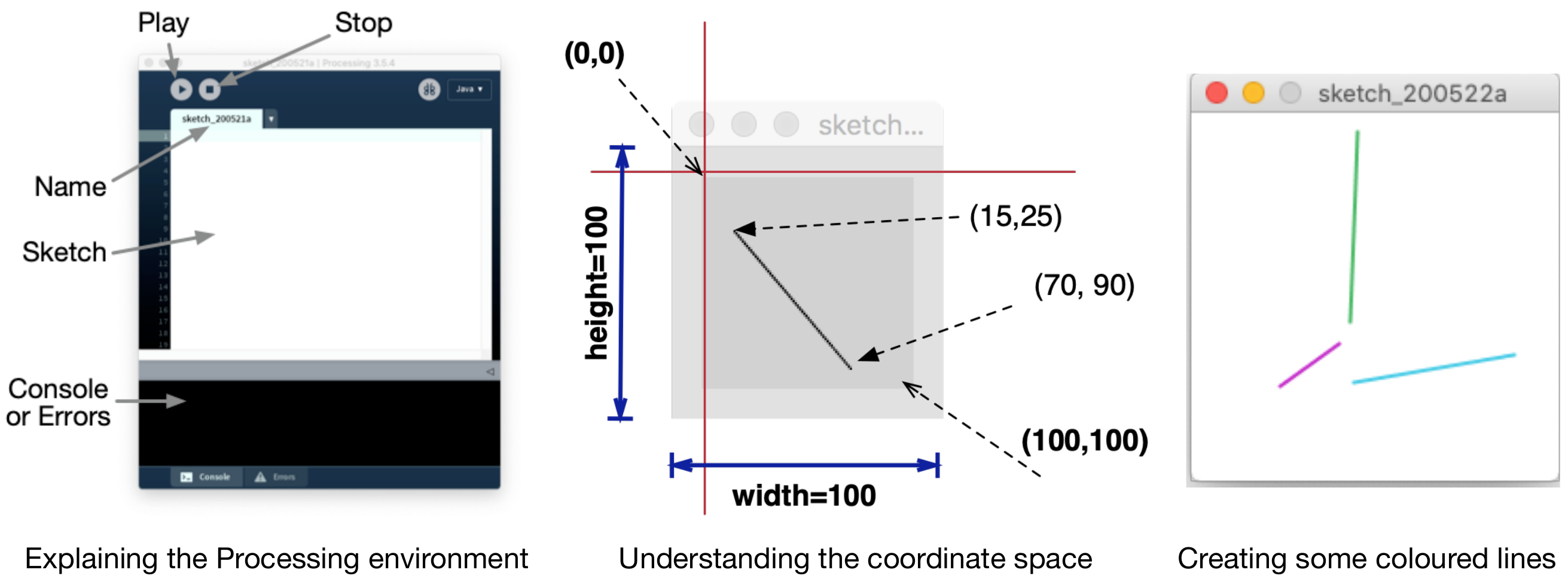}
    \caption{In the introduction to processing.org exercise we explain the Processing sketch environment, coordinate space, and get students to create random-positioned lines.}
    \label{fig:introprocessing}
\end{figure}

In the third activity we chose to focus on \textbf{laser lights}. In this activity we explained about lasers and how laser light travels. The activity allows people to bend laser light in water (Figure~\ref{fig:laser}). We list the materials (pint glass, washing liquid, laser pointer) and provide the instructions in a YouTube video. In the fourth activity we focus on image manipulation. The activity takes the learner through adding ``\textbf{a rainbow to an image in Photoshop}''. Our aims, for both the laser-light and Photoshop activities, were to attract a broader set of learners to the site. Perhaps people who were not interested in coding would find these activities interesting and while visiting Project Rainbow would try some of the coding activities. In fact, the final activity gets people to install the Processing environment and create some simple coloured lines, see Figure~\ref{fig:introprocessing}. At degree level, in our introduction to computer graphics course, we add much more detail. We discuss different coordinate spaces (right handed, and left handed coordinate systems) and different ways to express lines. For example, in processing
{\ttfamily line(x1,y1,x2,y2);} or path descriptions in SVG {\small\ttfamily <path d="M 100 100 L 300 100 z" stroke="blue"/>} to moving the turtle in Logo graphics {\ttfamily \mbox{forward} 100}. However, we decided to keep the activity focused on setting up the environment and including basic computer graphics concepts. At this stage we use  \textit{static sketches} in Processing~\cite{reas2007processing} which kept the programs simple because it creates a single image that does not include animation or interaction.

\begin{figure*}
\hspace*{0mm}
    \begin{subfigure}[t]{.3\textwidth}
    \includegraphics[height=7.2cm]{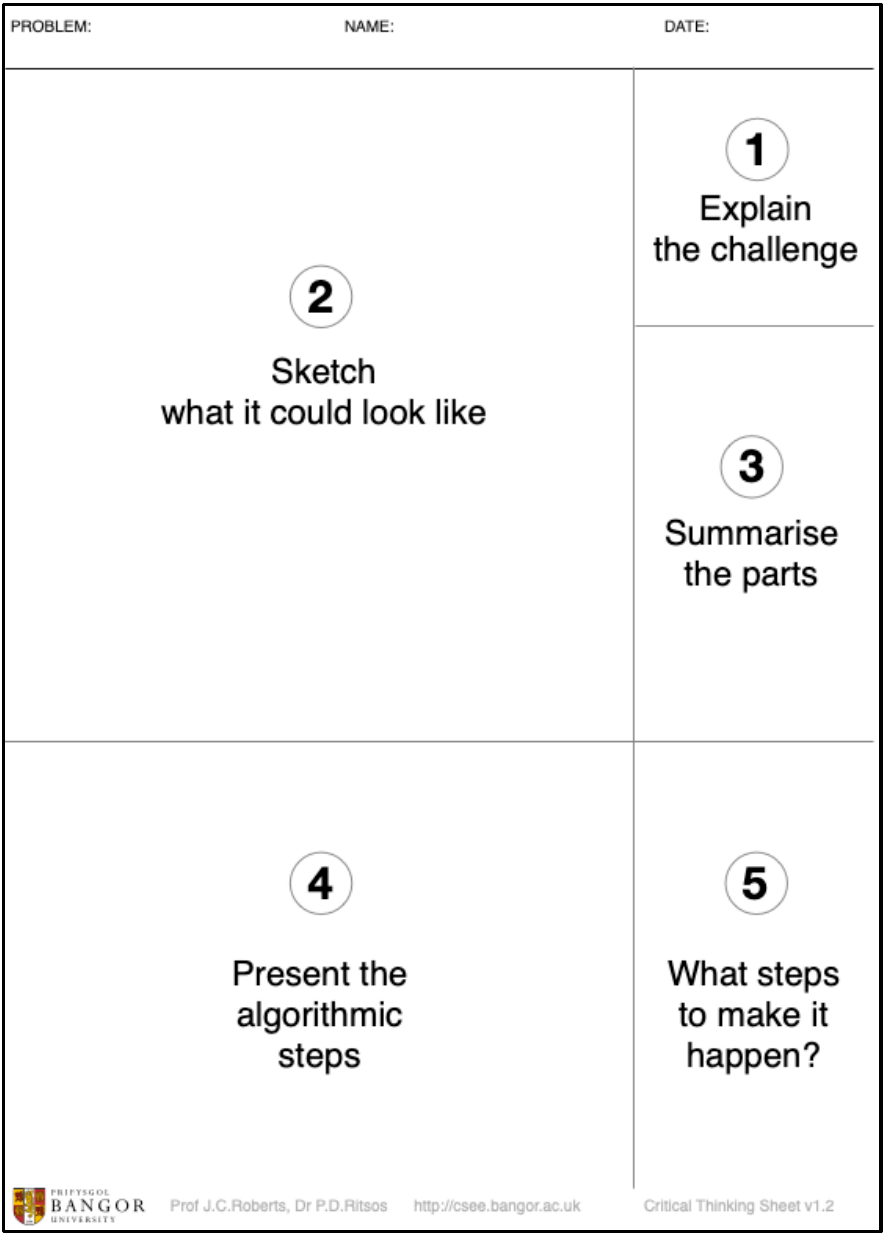}
    \caption{}
    \end{subfigure}\hspace{3mm}
    \begin{subfigure}[t]{.3\textwidth}
    \includegraphics[height=7.2cm]{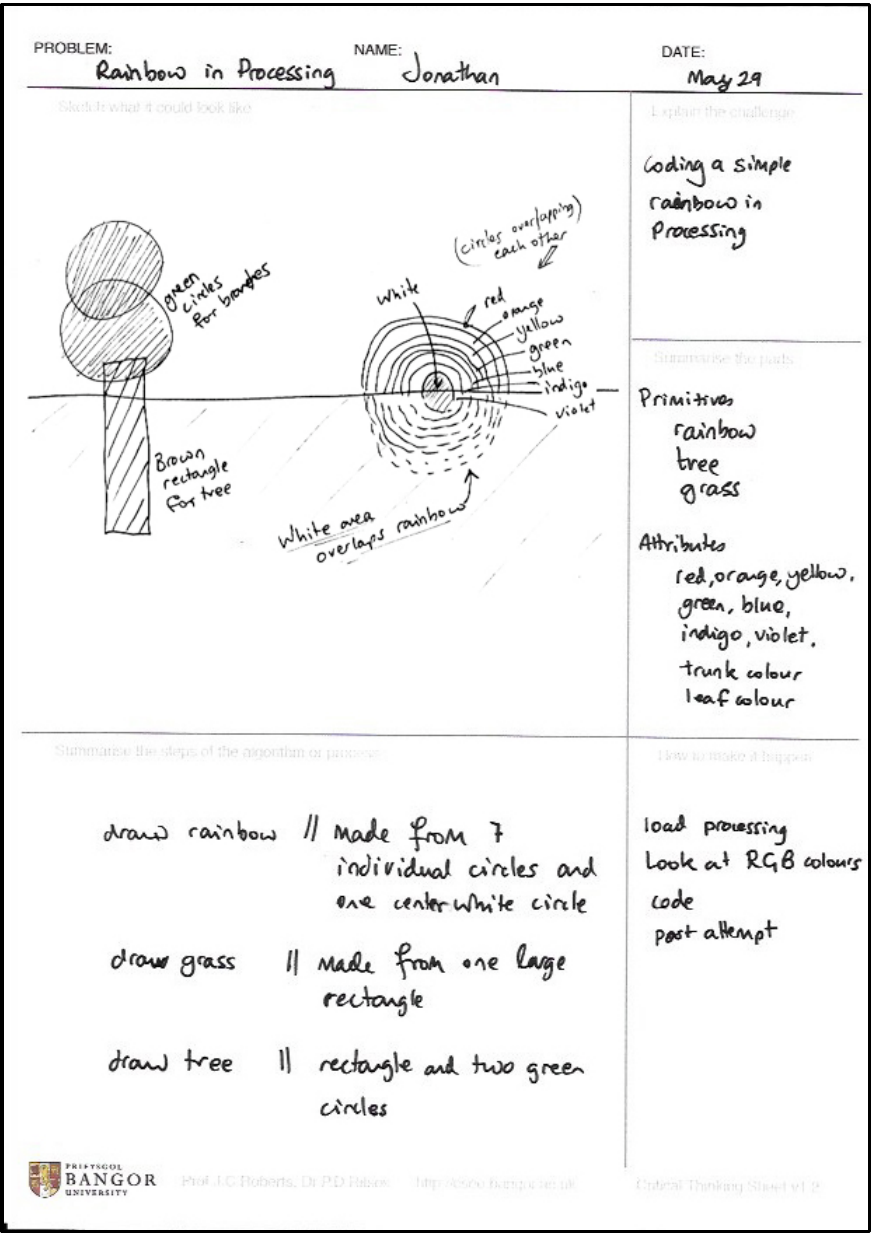}
    \caption{}
    \end{subfigure}\hspace{3mm}%
    \begin{subfigure}[t]{.3\textwidth}%
    \vspace{-7.15cm}\begin{minipage}[c]{8cm}%
    \includegraphics[height=2.5cm]{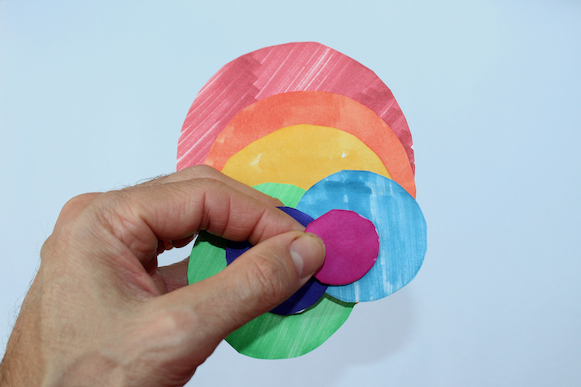}\\
    \includegraphics[height=2.cm]{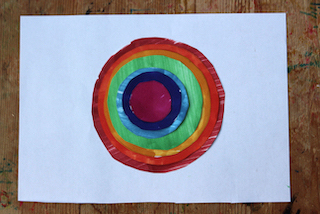}\hspace{2mm}\includegraphics[height=2.cm]{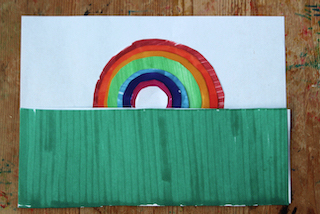}
    \includegraphics[height=2.5cm]{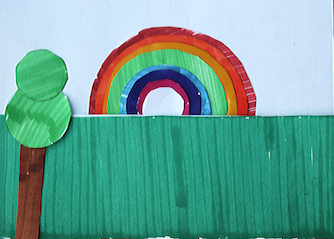}
    \end{minipage}
    \caption{}
    \end{subfigure}
    \caption{(a) The Critical Thinking Sheet consists of five panels, which students complete in turn~\cite{roberts2019critical,RobertsRitsos_CTS_2020}. (b) A completed CTS sheet for the rainbow scene example. (c) The physical mockup of the rainbow scene, used to explain the painters algorithm.\vspace{-3mm}}
    \label{fig:cts}
\end{figure*}

\section{\label{sec:critical} Planning activities -- critically thinking and sketching proposed solutions}
We believe that \textbf{critical thinking} is an important, yet sometimes, difficult skill to learn~\cite{alsaleh2020teaching}. We want learners to contemplate and reflect on their task and make plans and strategise appropriate solutions. We do not want learners to rush into coding~\cite{Roberts_ETAL2018_EVF,RobertsHeadleandRitsos15}. Instead we want them to critically think about their problem, make a sketch and write some ideas of how they could solve it. In this way, hopefully, learners will be able to quickly throw out bad ideas, and hence save time that they may have lost if they needed to restart the project (if they found their solution did not work). While the Five Design-Sheets could be suitable, we need a much quicker and focused methodology.

Our Critical thinking Sheet (CTS) method~\cite{roberts2019critical,RobertsRitsos_CTS_2020} gets learners to sketch a picture of their visual design, think through the main parts of the design, and consider the algorithmic steps to create the design. Figure~\ref{fig:cts}a shows the layout of the CTS sheet. Learners go through the panels in turn, and explain the challenge in panel (1), make a sketch of what the final output could look like in panel (2), summarise the main facets of the code in panel (3), present the main algorithmic steps that would be necessary to create the design in panel (4), and in panel (5) consider what they need to do next to implement it. An example of a completed sheet for the rainbow scene is shown in Figure~\ref{fig:cts}b. Following from lessons learnt with the Five Design-Sheet design-study~\cite{RobertsHeadleandRitsos15}, we encourage students to complete the CTS by hand. Students print the sheet, and sketch and write their solutions in pen or pencil. It is much easier for people to complete the sheets by hand and quicker for them to make sketches. If touch screens or other technologies are used, students can get distracted by technology rather than focusing on the main ideas and contemplating appropriate solutions. 

Sequentially completing the five steps of the CTS is important. When learners have to explain the challenge, and write it down in a small space, they are being forced to consider the main ideas. The act of thinking about appropriate summary words forces the person to think hard and critically about the ideas. Making a sketch helps learners externalise their vision. The picture, not only provides a record of the idea, but helps people to realise how close they are to completing the project. They can compare their solution in code to their sketched solution. When learners summarise the main `parts' or components of the system, they need to categorise the solutions by breaking it down into smaller `core' ideas. In this activity we encourage learners to write down a list of parts as \textit{primitives} and \textit{attributes}. We explain what each of these are and provide examples. In this activity we get the students to create a scene that has a simple rainbow, tree and grass. We make the scene from simple primitives of coloured circles and rectangles. Consequently, we can write the primitives as \textit{rainbow}, \textit{tree} and \textit{grass}; and attributes as \textit{rainbow} (red, orange, yellow, green, blue, indigo, violet), \textit{trunk colour} (brown) and \textit{leaf colour} (green). This is shown in the middle-right panel on the CTS in Figure~\ref{fig:cts}b.

To understand the graphics drawing process, we created a physical mockup of the scene, see Figure~\ref{fig:cts}c. This process will help the learner consider the steps needed to create the picture, using the Painters algorithm~\cite{foley1996computer}, where the rainbow parts are made from circles, they are positioned at the back. Subsequent circles partially occlude earlier circles. Then the grass is added, which occludes part of the coloured circles, allowing them to appear as semi-circles to make up the rainbow shape. Finally the tree is added, with the leaf circles on top, as they are the closest to the observer. Using this physical mockup, it is easy to explain how the visual illusion, turns the rainbow-coloured circles into arc-like visuals. We could have used arc primitives for each rainbow part, but this would not necessarily describe the painters algorithm.

\begin{figure}[ht]
    \centering
    \includegraphics[width=.8\columnwidth]{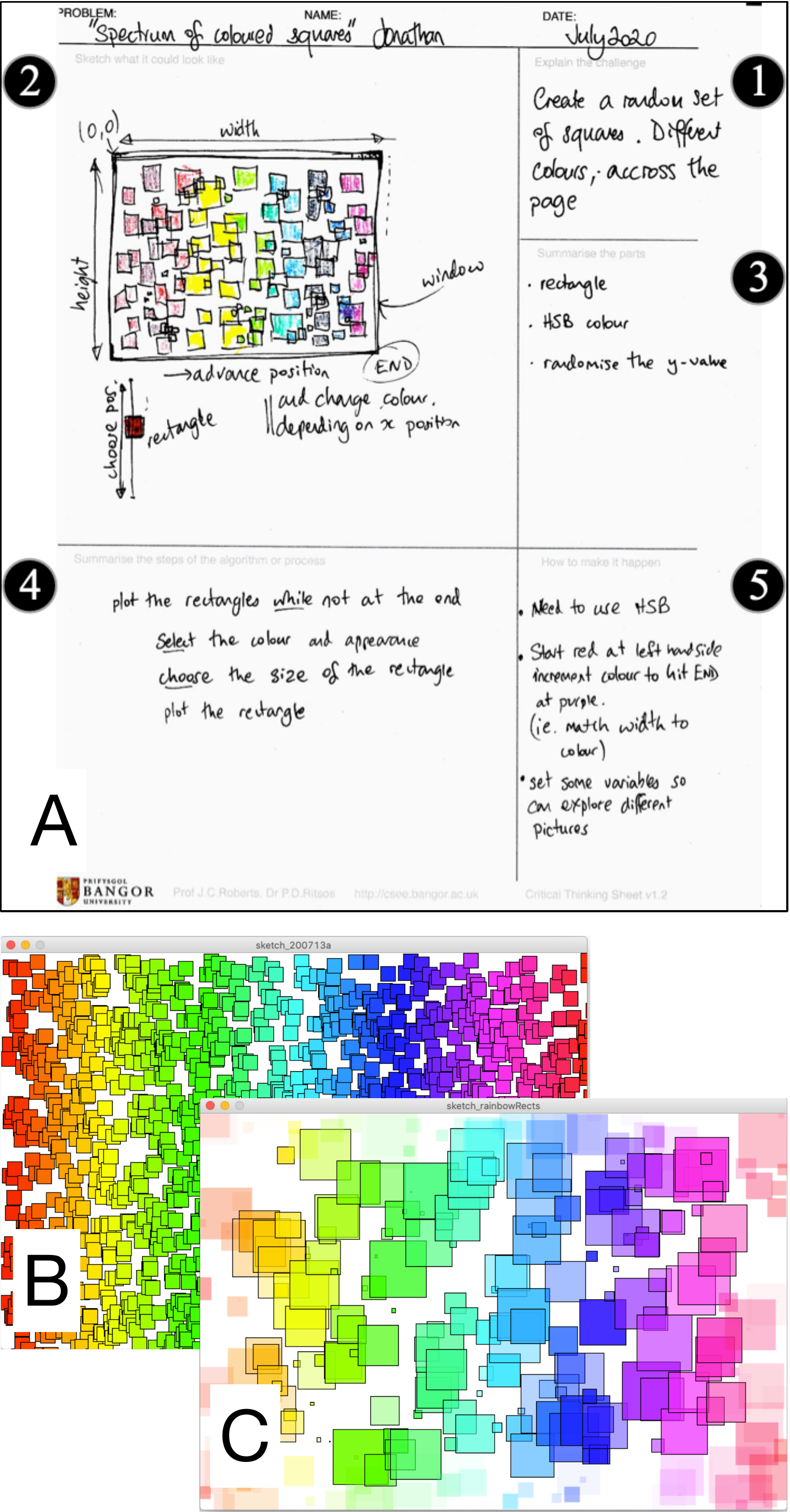}
    \caption{In this coding activity students create hundreds of rectangles that show a spectrum of colour. (A) students first perform a CTS study~\cite{roberts2019critical,RobertsRitsos_CTS_2020}, to create the basic code (B), which in a second activity they adapt and vary the size and appearance of the rectangles (C).}
    \label{fig:patterns}
    %\vspace{-2mm}
\end{figure}

\section{\label{sec:colour} Colour coding activities -- creating colourful patterns}
We developed a suite of activities that get learners creating  block patterns of coloured rectangles (see Figure~\ref{fig:patterns}). The first set of activities are created with a single \texttt{while} loop. To make the patterns more interesting we jitter the positions by adding or subtracting random numbers. Each of the activities get the student to create a CTS sheet for the activity, think through the loop structure (\texttt{while} or \texttt{for} loop), control the \texttt{ColorMode()} to define RGB or HSV colour spaces, and finally adapt the code using their own creativity guided by a series of additional tasks.
\begin{figure}
    \centering
    \includegraphics[width=\columnwidth]{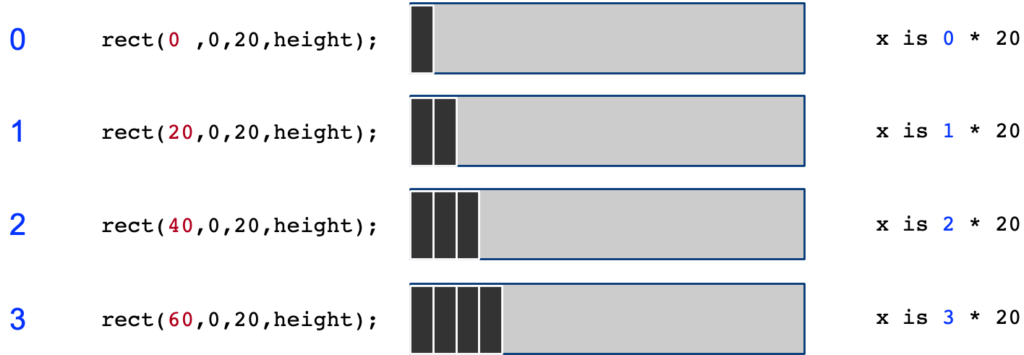}
    \caption{We explain some best practices in computer graphics; in this example we show how to use a for-loop to index position, and multiply by the object width/height to locate the rectangle.}
    \label{fig:loops}
\end{figure}

The second set of colour coding activities use two for loops to create gridded patterns. Using this strategy, students have some activities on simple loop structures, and then start thinking about plotting \texttt{xy} coordinates using two \texttt{for} loops. We wish to explain some best practices of plotting parts on Cartesian grid. Figure~\ref{fig:loops} shows one of our  visual explanations of the loops, where we use the \texttt{for} loops to control the grid location, and multiply against the width (or height) of the rectangle to calculate the exact location in screen coordinates.

\section{\label{sec:advanced} Advanced activities -- augmented reality and further learning}
Our goals with the advanced set of activities were to cover subjects that learners may not have done, and to focus on areas that were associated with computer graphics research that is being undertaken by researchers at our University~\cite{roberts2021visualisation}. For instance, we developed some rainbow activities in WebXR using A-Frame and also Augmented reality rainbows. Colleagues have been researching methods to move beyond traditional desktop interfaces~\cite{Roberts-et-al-CGA2014}, and develop interfaces to help people become immersed in their data~\cite{Butcher-et-al-TVCG-2021}. In other work, researchers have been creating novel visualisation tools, such as to visualise and explain paths of redress in law and administrative justice~\cite{RobertsETAL2021ARTEMUS}, visualising student journeys~\cite{Gray-et-al-CAEH-2020} or multiple views~\cite{ManeeaRobertsTool2020,ChenETAL2021,RobertsETALMultipleViewsMeanings2019}. Subsequently we wrote an activity using the D3.js library, which is a JavaScript library to help developers manipulate documents based on data~\cite{BostockETALD3-2011}.  The D3.js activity was used to develop code to `sing' the colours of the rainbow. Finally, we created an activity, which led learners through steps to create a rainbow in pure CSS code.

\begin{figure}
    \centering
    \includegraphics[width=\columnwidth]{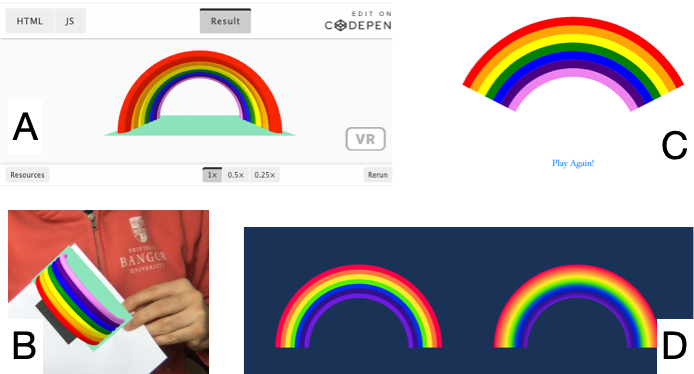}
    \caption{A selection of the advanced activities. (A) shows the A-frame rainbow, (B) augmented reality demonstration, (C) D3.js sing a rainbow that animates the colours through the words of the song, and (D) CSS rainbows.}
    \label{fig:advanced}
\end{figure}

Each of these activities followed a similar structure: we explained the goal of the activity and the vision of what they were going to create, ran through the steps to achieve the goal, and then get the student to experiment and adapt the code. Figure~\ref{fig:advanced} shows screenshots of the final output. Both the D3.js and CSS activities explained how to do animation in these systems, and we introduced external systems such as CodePen.io, which enables people to edit and run the code in the browser.

\section{\label{sec:end} Discussion and conclusions}
While the coronavirus pandemic has been terrible, where many people have lost their lives and lively-hoods permanently changed, it did motivate us to come together as a team of graphics and visualisation experts and share some of our knowledge with the world. Developing Project Rainbow has been a rewarding experience, and we have had some positive feedback from participants saying thanks, and expressing that they have found them interesting and rewarding to follow. It is interesting to look at the statistics on the site. Some activities have been more popular; the activities on creating a coloured grid in Processing, pure CSS rainbows, a rainbow in code, and learning Hue Saturation and Value (HSB) were the most popular. With participants coming from the top five countries of Australia, United States, UK, India and Canada, respectively. Most referrals came from Facebook and Twitter. This seems sensible, because they were our two primary sources of advertisement. We have re-used the materials in several classes and workshops, especially we used the Welsh versions of the ten processing.org programming activities in the National Eisteddfod of Wales.

One of the main challenges we found is that it takes much effort to create a set of activities that are publicly available. Academics and teachers are busy people doing teaching and research. Much care and attention is needed to guarantee that the information is correct, clear and can be followed by people. This takes a huge amount of time. We found that we needed to include many examples, pictures and code segments, to make the information clear and easy to follow. Each of these images needed to be designed, created, saved as images, and uploaded to the WordPress media store, labelled with the alternative text and included in the post. Like any publishing, developing the content of each activity also takes time and effort. Many activities were refined, edited, and reorganised many times, before they were finally checked by a colleague, and before they were translated into Welsh. In addition, while we now have an excellent bilingual resource, it certainly takes more effort to write the activities in two languages. However, the tools and processes that we used, especially the WordPress translation plugin, were extremely useful to make this process easier. 

We believe that the time we spent at the start of the project, in carefully discussing and planning how to setup the WordPress site helped us to save time in the end. Especially developing the set of guidelines of how to structure and write the post, helped authors write the activities and structure them in a consistent way. The structure, now created, can be re-used in other projects. Importantly, by developing guidelines over selecting and allocating the \textit{categories} (beginner, intermediate and advanced), and the additional \textit{tag} words, enabled us to use the search and organisation facilities of WordPress in a consistent way. It also helps to make the site more consistent and simple to use.

We set out to create a set of activities focused on colour and rainbows; we have achieved these goals, and have created a structure and ideology that other people can follow. Furthermore, at the start of the project we were keen to develop learning-activities, with the knowledge that they were useful. On reflection, we now have a stronger view, that learning \textit{activities} are necessary. We have since, taken ideas that we have learnt in this project and applied them to our own degree teaching. For instance, we re-organised the material of  a creative computing, storytelling and design module, into the research, design, development and reflection structure. Each week was focuced around a set of activities. We developed videos of the activities, wrote guidance material, and set a series of tasks for the students to perform.

Project Rainbow has been a success. It has enabled us (as academics) to create a range of public facing activities. Members of the public, and those at the National Welsh Eisteddfod have followed them in both Welsh and English, and we have learnt and applied the structure to other projects and courses. We hope that other researchers and teachers take up the ideas and learn and apply the structures that we have put in place to their own projects.

\section*{Acknowledgements}
We acknowledge the staff of the School of Computer Science and Electronic Engineering for supporting this project. In particular we acknowledge the many colleagues who have helped to deliver the material, without which this project would not have been completed. Dr Panagiotis Ritsos for the virtual reality and augmented reality activities, Dr Peter Butcher for the CSS activity, Dr Daniel Roberts for the laser activity and Welsh translations, Mr Joe Mearman for help with the website design, and Dr Cameron Gray for help with setting up the website, web policies and writing the D3.js activity.
%-------------------------------------------------------------------------
% bibtex
\bibliographystyle{eg-alpha-doi} 
\bibliography{rainbow.bib}       

\newcommand{\etalchar}[1]{$^{#1}$}
\begin{thebibliography}{\uppercase{RAmB{\etalchar{*}}19}}

\bibitem[Als20]{alsaleh2020teaching}
\textsc{Alsaleh N.~J.}:
\newblock Teaching critical thinking skills: Literature review.
\newblock \emph{Turkish Online Journal of Educational Technology-TOJET 19}, 1
  (2020), 21--39.

\bibitem[AMR20]{ManeeaRobertsTool2020}
\textsc{Al-Maneea H., Roberts J.}:
\newblock {A tool to help lay out Multiple View Visualisations guided by view
  analysis}.
\newblock In \emph{Posters EuroVis 2020, 25-29 May 2020} (May 2020).

\bibitem[BE91]{bonwell1991active}
\textsc{Bonwell C.~C., Eison J.~A.}:
\newblock \emph{Active Learning: Creating Excitement in the Classroom. 1991
  ASHE-ERIC Higher Education Reports.}
\newblock ERIC, 1991.

\bibitem[BGP{\etalchar{*}}11]{borkin2011}
\textsc{Borkin M., Gajos K., Peters A., Mitsouras D., Melchionna S., Rybicki
  F., Feldman C., Pfister H.}:
\newblock Evaluation of artery visualizations for heart disease diagnosis.
\newblock \emph{IEEE Transactions on Visualization and Computer Graphics 17},
  12 (2011), 2479--2488.
\newblock \href {https://doi.org/10.1109/TVCG.2011.192}
  {\path{doi:10.1109/TVCG.2011.192}}.

\bibitem[BJR21]{Butcher-et-al-TVCG-2021}
\textsc{Butcher P.~W., John N.~W., Ritsos P.~D.}:
\newblock {VRIA: A Web-based Framework for Creating Immersive Analytics
  Experiences}.
\newblock \emph{IEEE Transactions on Visualization and Computer Graphics 27},
  07 (July 2021), 3213--3225.
\newblock \href {https://doi.org/10.1109/TVCG.2020.2965109}
  {\path{doi:10.1109/TVCG.2020.2965109}}.

\bibitem[BOH11]{BostockETALD3-2011}
\textsc{Bostock M., Ogievetsky V., Heer J.}:
\newblock {D$^{3}$ Data-Driven Documents}.
\newblock \emph{IEEE Transactions on Visualization and Computer Graphics 17},
  12 (2011), 2301--2309.
\newblock \href {https://doi.org/10.1109/TVCG.2011.185}
  {\path{doi:10.1109/TVCG.2011.185}}.

\bibitem[BTI07]{borland2007rainbow}
\textsc{Borland D., Taylor~Ii R.~M.}:
\newblock Rainbow color map (still) considered harmful.
\newblock \emph{IEEE Computer Graphics and Applications 27}, 2 (2007), 14--17.
\newblock \href {https://doi.org/10.1109/MCG.2007.323435}
  {\path{doi:10.1109/MCG.2007.323435}}.

\bibitem[Bux10]{buxton2010sketching}
\textsc{Buxton B.}:
\newblock \emph{Sketching user experiences: getting the design right and the
  right design}.
\newblock Morgan Kaufmann, 2010.

\bibitem[BV18]{bell2018cs}
\textsc{Bell T., Vahrenhold J.}:
\newblock {CS} unplugged—how is it used, and does it work?
\newblock In \emph{Adventures between lower bounds and higher altitudes}.
  Springer, 2018, pp.~497--521.

\bibitem[CSH20]{CrameriETAL2020_MisuseOfColour}
\textsc{Crameri F., Shephard G., Heron P.}:
\newblock The misuse of colour in science communication.
\newblock \emph{Nature Communications 11}, 5444 (2020).
\newblock \href {https://doi.org/10.1038/s41467-020-19160-7}
  {\path{doi:10.1038/s41467-020-19160-7}}.

\bibitem[CZL{\etalchar{*}}21]{ChenETAL2021}
\textsc{Chen X., Zeng W., Lin Y., AI-maneea H.~M., Roberts J., Chang R.}:
\newblock Composition and configuration patterns in multiple-view
  visualizations.
\newblock \emph{IEEE Transactions on Visualization and Computer Graphics 27}, 2
  (2021), 1514--1524.
\newblock \href {https://doi.org/10.1109/TVCG.2020.3030338}
  {\path{doi:10.1109/TVCG.2020.3030338}}.

\bibitem[FVVD{\etalchar{*}}96]{foley1996computer}
\textsc{Foley J.~D., Van F.~D., Van~Dam A., Feiner S.~K., Hughes J.~F., Angel
  E., Hughes J.}:
\newblock \emph{Computer graphics: principles and practice}, vol.~12110.
\newblock Addison-Wesley Professional, 1996.

\bibitem[GPR20]{Gray-et-al-CAEH-2020}
\textsc{Gray C.~C., Perkins D., Ritsos P.~D.}:
\newblock {Degree Pictures: Visualizing the university student journey}.
\newblock \emph{Assessment \& Evaluation in Higher Education 20}, 4 (Aug.
  2020), 568--578.
\newblock \href {https://doi.org/10.1080/02602938.2019.1676397}
  {\path{doi:10.1080/02602938.2019.1676397}}.

\bibitem[HA17]{HeAdarVisitCards}
\textsc{He S., Adar E.}:
\newblock Vizitcards: A card-based toolkit for infovis design education.
\newblock \emph{IEEE Transactions on Visualization and Computer Graphics 23}, 1
  (Jan 2017), 561--570.
\newblock \href {https://doi.org/10.1109/TVCG.2016.2599338}
  {\path{doi:10.1109/TVCG.2016.2599338}}.

\bibitem[HCT{\etalchar{*}}14]{huron2014constructiveviz}
\textsc{Huron S., Carpendale S., Thudt A., Tang A., Mauerer M.}:
\newblock Constructive visualization.
\newblock In \emph{Proceedings of the ACM conference on Designing Interactive
  Systems (DIS)} (2014), ACM, pp.~433--442.
\newblock \href {https://doi.org/10.1145/2598784.2598806}
  {\path{doi:10.1145/2598784.2598806}}.

\bibitem[Ken99]{kennedy1999role}
\textsc{Kennedy M.~M.}:
\newblock The role of preservice teacher education.
\newblock \emph{Teaching as the learning profession: Handbook of policy and
  practice} (1999), 54--85.

\bibitem[Kep14]{keppell2014personalised}
\textsc{Keppell M.}:
\newblock Personalised learning strategies for higher education.
\newblock In \emph{The Future of Learning and Teaching in Next Generation
  Learning Spaces (International Perspectives on Higher Education Research,
  Vol. 12)}. Emerald Group Publishing Limited, 2014, pp.~3--21.
\newblock \href {https://doi.org/10.1108/S1479-362820140000012001}
  {\path{doi:10.1108/S1479-362820140000012001}}.

\bibitem[KK05]{KolbKolb2005}
\textsc{Kolb A.~Y., Kolb D.~A.}:
\newblock Learning styles and learning spaces: Enhancing experiential learning
  in higher education.
\newblock \emph{Academy of Management Learning \& Education 4}, 2 (2005),
  193--212.

\bibitem[Kra02]{krathwohl2002revision}
\textsc{Krathwohl D.~R.}:
\newblock A revision of bloom's taxonomy: An overview.
\newblock \emph{Theory Into Practice 41}, 4 (2002), 212--218.
\newblock \href {https://doi.org/10.1207/s15430421tip4104\_2}
  {\path{doi:10.1207/s15430421tip4104\_2}}.

\bibitem[L\"16]{Lowgren2016}
\textsc{L\"{o}wgren J.}:
\newblock On the significance of making in interaction design research.
\newblock \emph{Interactions 23}, 3 (Apr. 2016), 26--33.
\newblock \href {https://doi.org/10.1145/2904376} {\path{doi:10.1145/2904376}}.

\bibitem[MMAM14]{McKennaETAL2014}
\textsc{McKenna S., Mazur D., Agutter J., Meyer M.}:
\newblock Design activity framework for visualization design.
\newblock \emph{IEEE Transactions on Visualization and Computer Graphics 20},
  12 (Dec 2014), 2191--2200.
\newblock \href {https://doi.org/10.1109/TVCG.2014.2346331}
  {\path{doi:10.1109/TVCG.2014.2346331}}.

\bibitem[Mun09]{Munzner2009}
\textsc{Munzner T.}:
\newblock A nested process model for visualization design and validation.
\newblock \emph{{IEEE} Transactions on Visualization and Computer Graphics 15}
  (Nov 2009), 921--928.
\newblock \href {https://doi.org/10.1109/TVCG.2009.111}
  {\path{doi:10.1109/TVCG.2009.111}}.

\bibitem[RAmB{\etalchar{*}}19]{RobertsETALMultipleViewsMeanings2019}
\textsc{Roberts J.~C., Al-maneea H., Butcher P. W.~S., Lew R., Rees G., Sharma
  N., Frankenberg-Garcia A.}:
\newblock Multiple views: different meanings and collocated words.
\newblock \emph{Computer Graphics Forum 38}, 3 (2019), 79--93.
\newblock \href {https://doi.org/https://doi.org/10.1111/cgf.13673}
  {\path{doi:https://doi.org/10.1111/cgf.13673}}.

\bibitem[RBSN21]{RobertsETAL2021ARTEMUS}
\textsc{Roberts J.~C., Butcher P., Sherlock A., Nason S.}:
\newblock Explanatory journeys: Visualising to understand and explain
  administrative justice paths of redress.
\newblock \emph{IEEE Transactions on Visualization and Computer Graphics}
  (2021).
\newblock Accepted/In press.
\newblock URL: \url{http://arxiv.org/abs/2107.14013}.

\bibitem[RF07]{reas2007processing}
\textsc{Reas C., Fry B.}:
\newblock \emph{Processing: a programming handbook for visual designers and
  artists}.
\newblock Mit Press, 2007.

\bibitem[RHR16]{RobertsHeadleandRitsos15}
\textsc{Roberts J.~C., Headleand C., Ritsos P.~D.}:
\newblock Sketching designs using the five design-sheet methodology.
\newblock \emph{IEEE Transactions on Visualization and Computer Graphics 22}, 1
  (Jan 2016), 419--428.
\newblock \href {https://doi.org/10.1109/TVCG.2015.2467271}
  {\path{doi:10.1109/TVCG.2015.2467271}}.

\bibitem[RHR17]{RobertsHeadleandRitsos2017}
\textsc{Roberts J.~C., Headleand C.~J., Ritsos P.~D.}:
\newblock \emph{Five Design-Sheets -- Creative design and sketching in
  Computing and Visualization}.
\newblock Springer International Publishing, 2017.
\newblock \href {https://doi.org/10.1007/978-3-319-55627-7}
  {\path{doi:10.1007/978-3-319-55627-7}}.

\bibitem[RK01]{RogowitzKalvin2001_WhichBlair}
\textsc{Rogowitz B., Kalvin A.}:
\newblock The ``which blair project'': a quick visual method for evaluating
  perceptual color maps.
\newblock In \emph{Proceedings Visualization, 2001. VIS'01} (2001),
  pp.~183--556.
\newblock \href {https://doi.org/10.1109/VISUAL.2001.964510}
  {\path{doi:10.1109/VISUAL.2001.964510}}.

\bibitem[Rob21]{ProjectRainbow2021}
\textsc{Roberts J.~C.}:
\newblock {Project Rainbow -- STEM} activities focusing on colours, light and
  optical effects., 2021.
\newblock URL: \url{https://csee.bangor.ac.uk/}.

\bibitem[RR19]{roberts2019critical}
\textsc{Roberts J.~C., Ritsos P.~D.}:
\newblock Critical thinking sheets: Encouraging critical thought and sketched
  implementation design.
\newblock In \emph{EduCHI 2019 Symposium: Global Perspectives on HCI Education,
  CHI Conference on Human Factors in Computing Systems (ACM CHI 2019)} (2019).

\bibitem[RR20]{RobertsRitsos_CTS_2020}
\textsc{Roberts J.~C., Ritsos P.~D.}:
\newblock {Critical Thinking Sheet (CTS) for Design Thinking in Programming
  Courses}.
\newblock In \emph{Eurographics 2020 - Education Papers} (2020), Romero M.,
  Sousa~Santos B., (Eds.), The Eurographics Association, pp.~17--23.
\newblock \href {https://doi.org/10.2312/eged.20201029}
  {\path{doi:10.2312/eged.20201029}}.

\bibitem[RRB{\etalchar{*}}14]{Roberts-et-al-CGA2014}
\textsc{Roberts J.~C., Ritsos P.~D., Badam S.~K., Brodbeck D., Kennedy J.,
  Elmqvist N.}:
\newblock {Visualization Beyond the Desktop - the next big thing}.
\newblock \emph{IEEE Computer Graphics and Applications 34}, 6 (Nov. 2014),
  26--34.
\newblock \href {https://doi.org/10.1109/MCG.2014.82}
  {\path{doi:10.1109/MCG.2014.82}}.

\bibitem[RRH17]{Roberts-et-al-PDVW-2017}
\textsc{Roberts J.~C., Ritsos P.~D., Headleand C.}:
\newblock {Experience and Guidance for the use of Sketching and low-fidelity
  Visualisation-design in teaching}.
\newblock In \emph{Pedagogy of Data Visualization Workshop, IEEE Conference on
  Visualization (VIS), Phoenix, Arizona, USA} (Oct. 2017), Joshi A., Adar E.,
  Bertini E., Engle S., Hearst M., Keefe D., (Eds.).

\bibitem[RRJH18]{Roberts_ETAL2018_EVF}
\textsc{{Roberts} J.~C., {Ritsos} P.~D., {Jackson} J.~R., {Headleand} C.}:
\newblock The explanatory visualization framework: An active learning framework
  for teaching creative computing using explanatory visualizations.
\newblock \emph{IEEE Transactions on Visualization and Computer Graphics 24}, 1
  (2018), 791--801.
\newblock \href {https://doi.org/10.1109/TVCG.2017.2745878}
  {\path{doi:10.1109/TVCG.2017.2745878}}.

\bibitem[RRK{\etalchar{*}}21]{roberts2021visualisation}
\textsc{Roberts J.~C., Ritsos P.~D., Kuncheva L., Vidal F., Lim I.~S.,
  Ap~Cenydd L., Teahan W., Mansoor S., Gray C., Perkins D.}:
\newblock {Visualisation Data Modelling Graphics (VDMG) at Bangor}.
\newblock In \emph{Research labs and projects: Eurographics 2021: the 42nd
  Annual Conference of the European Association for Computer Graphics} (2021).

\bibitem[RSDB20]{Ridley_ETAL_VisToolsActivity2020}
\textsc{Ridley A., Schöttler S., Dadzie A.-S., Bach B.}:
\newblock The vistools marketplace: An activity to understand the landscape of
  visualisation tools.
\newblock In \emph{VisActivities: IEEE VIS Workshop on Data Vis Activities to
  Facilitate Learning, Reflecting, Discussing, and Designing, held in
  conjunction with IEEE VIS 2020, Salt Lake City, UT.} (October 2020).
\newblock URL: \url{visactivities.github.io.}

\bibitem[RT98]{RogowitzTreinish1998}
\textsc{Rogowitz B., Treinish L.}:
\newblock Data visualization: the end of the rainbow.
\newblock \emph{IEEE Spectrum 35}, 12 (1998), 52--59.
\newblock \href {https://doi.org/10.1109/6.736450}
  {\path{doi:10.1109/6.736450}}.

\bibitem[Sim73]{Simon1973structure}
\textsc{Simon H.~A.}:
\newblock The structure of ill structured problems.
\newblock \emph{Artificial Intelligence 4}, 3-4 (1973), 181--201.

\bibitem[SMM12]{Sed2012a}
\textsc{Sedlmair M., Meyer M., Munzner T.}:
\newblock Design study methodology: Reflections from the trenches and the
  stacks.
\newblock \emph{{IEEE} Transactions on Visualization and Computer Graphics 18},
  12 (2012), 2431--2440.
\newblock \href {https://doi.org/10.1109/TVCG.2012.213}
  {\path{doi:10.1109/TVCG.2012.213}}.

\bibitem[Tob00]{tobin2000all}
\textsc{Tobin D.~R.}:
\newblock \emph{All learning is self-directed: How organizations can support
  and encourage independent learning}.
\newblock American Society for Training and Development, 2000.

\bibitem[War12]{Ware2012}
\textsc{Ware C.}:
\newblock \emph{Information Visualization: Perception for Design}, 3~ed.
\newblock Morgan Kaufmann Series in Interactive Technologies. Morgan Kaufmann,
  Amsterdam, 2012.
\newblock URL: \url{http://www.sciencedirect.com/science/book/9780123814647}.

\end{thebibliography}

% biblatex with biber
% \printbibliography                
\end{document}